\documentclass[aps,prl,twocolumn,showpacs,superscriptaddress]{revtex4}
\usepackage{graphicx}
\usepackage{amsmath}
\usepackage{amssymb}
\usepackage{epstopdf}
\usepackage{amsfonts}
\usepackage{dcolumn}
\begin{document}
\title{Harmonically trapped Fermi gas: Temperature dependence of the
Tan contact}
\author{Yangqian Yan}
\affiliation{Department of Physics and Astronomy,
Washington State University,
  Pullman, Washington 99164-2814, USA}
\author{D. Blume}
\affiliation{Department of Physics and Astronomy,
Washington State University,
  Pullman, Washington 99164-2814, USA}
\affiliation{ITAMP, Harvard-Smithsonian Center for Astrophysics,
Cambridge, Massachusetts 02138, USA}
\date{\today}

\begin{abstract}
Ultracold atomic gases with short-range interactions 
are characterized by a number of universal species-independent 
relations.
Many of these relations involve the two-body Tan contact.
Employing the canonical ensemble, we determine the
Tan contact for small harmonically trapped 
two-component Fermi gases at unitarity over a wide range of
temperatures, including the zero and high temperature regimes.
A cluster expansion that describes the properties of the
$N$-particle system in terms of those of smaller subsystems is introduced 
and shown to provide an accurate description of the contact
in the high temperature regime.
Finite-range corrections are quantified and the role
of the Fermi statistics is elucidated by comparing
results for Fermi, Bose and Boltzmann statistics.
\end{abstract}
\maketitle

Systems with short-range interactions are characterized by universal
relations that are independent of the details of the underlying
interactions.
The Tan contact~\cite{Tan1,Tan2,Tan3,castin,braaten}, 
e.g., enters into a large number of universal
relations and connects physically distinct quantities such as
the large momentum tail,
the inelastic loss rate, the number of pairs with small interparticle
distances, and certain characteristics of
radio frequency (rf) spectra.
A striking feature of many of the universal relations is that
they apply to homogeneous {\em{and}} inhomogeneous systems at
zero {\em{and}} finite temperature.

Yet, although many universal relations that evolve around the 
Tan contact are 
known,
only a few explicit 
measurements~\cite{earlyJin,Vale2010,Vale2011,Jin12,hoinka} and 
calculations~\cite{DailyBlume,werner09,strinati,drut,svistunov,boettcher,zwerger,doggen} 
of 
the Tan contact exist.
At present, the dependence of the Tan contact of the
spin-balanced homogeneous two-component Fermi gas at unitarity
on the temperature is highly debated.
While experimental data~\cite{Jin12}, 
extracted from the rf lineshape
of a potassium mixture in two different hyperfine states,
indicate a fairly sharp jump of the contact around the
superfluid 
transition temperature, two sets of Monte
Carlo simulations~\cite{drut,svistunov}---which 
disagree with each other---do
not seem to see this jump.

This paper considers small Fermi gases 
consisting of $N_1$ spin-up and $N_2$ spin-down
fermions under external spherically symmetric harmonic
confinement.
Working in the canonical ensemble,
we determine the Tan contact at unitarity 
with an accuracy at the few percent level as a 
function of the temperature, including the low (near zero) temperature 
regime and the high temperature regime.
Our main findings are:
{\em{(i)}}
The high temperature tail of the contact
$C_{N_1,N_2}$ is well approximated by the contacts of the 
$(N_1',N_2')$ subclusters with
$N_1'+N_2' < N$, where $N_1' \le N_1$ 
and $N_2' \le N_2$. 
The connection between this cluster 
expansion, which can be applied to 
any thermodynamic observable calculated in the canonical ensemble, and the 
virial expansion that has recently been applied extensively to determine
thermodynamic properties of atomic gases in the grand canonical 
ensemble~\cite{JasonHo,Drummond,liurev},
is described.
{\em{(ii)}}
While the contact of the trapped $(N_1,N_2)=(1,1)$
and $(2,2)$
systems is maximal at $T=0$,
that of the 
$(2,1)$, $(3,1)$ and $(4,1)$ systems shows a maximum
at finite temperature.
A microscopic interpretation of this behavior is offered.
{\em{(iii)}}
For the cases studied, the contact shows
a non-negligible
dependence on the range $r_0$
of the underlying two-body potential at low
temperature; in the high temperature regime,
in contrast, the range dependence is negligible.
{\em{(iv)}}
Fermi statistics plays a role at temperatures where three-body physics
is non-negligible.
The role of the Fermi statistics is elucidated by turning
the exchange symmetry off and by switching to Bose statistics.

The two-component Fermi gas
consisting of $N$ atoms with mass $m$ 
and position vectors $\mathbf{r}_j$
($j=1,\cdots,N$)
is described by the model Hamiltonian $H$,
\begin{equation}\label{general Hamiltonian}
  H=\sum_{j=1}^{N}\left(\frac{-\hbar^2}{2m}\nabla_j^2+\frac{1}{2}m
  \omega^2\mathbf{r}_{j}^2
\right)+
\sum_{j=1}^{N_1} \sum_{k=N_1+1}^{N}
V_{\text{tb}}(\mathbf{r}_{jk}),
\end{equation} 
where
$\omega$ denotes the angular trapping frequency.
We consider two different interspecies
two-body potentials $V_{\text{tb}}$,
a 
regularized zero-range
pseudopotential $V_{\text{F}}$~\cite{Yang1957}
and a short-range Gaussian potential
$V_{\text{G}}$ with depth $U_0$ ($U_0<0$) and
range $r_0$,
$V_{\text{G}}(\mathbf{r}_{jk})=
U_{0} \exp 
[ -
\mathbf{r}_{jk}^2/(2r_{0}^2)
]$.
For a given $r_0$, we adjust 
$U_0$ such that $V_{\text{G}}$ supports a single zero-energy
$s$-wave bound state in free space, i.e., such that the
$s$-wave scattering length $a_s$ diverges.
Our calculations use $r_0 \ll a_{\text{ho}}$,
where $a_{\text{ho}}$ is the harmonic
oscillator length, $a_{\text{ho}}=\sqrt{\hbar/(m \omega)}$.
This Letter considers temperatures ranging from
$T=0$ to $k_B T \gg E_{\text{ho}}$, where $E_{\text{ho}} = \hbar \omega$.
The largest temperatures considered are chosen such that the
two-body interactions can be reliably parametrized by the $s$-wave scattering
length and corresponding effective range correction, i.e., so that
higher partial wave contributions in the two-body sector can be neglected.

To determine the Tan contact $C_{N_1,N_2}$, we employ the adiabatic
and pair relations,
\begin{equation} 
  C_{N_1,N_2}=\frac{4\pi m}{\hbar^2}
  \left\langle \frac{\partial E(a_s)}{\partial(-a_s^{-1})} 
\right\rangle_{\text{th}}
  \label{eq_adiabaticrelation}
\end{equation}
and
\begin{equation}
  C_{N_1,N_2}= 4 \pi \lim_{s\to0}
\frac{ \langle N_{\text{pair}}^{r<s} \rangle_{\text{th}}}{s};
  \label{eq_pairrelation}
\end{equation}
here, the $\langle \cdot \rangle_{\text{th}}$ 
notation indicates a thermal average
and $E(a_s)$ denotes the energy of the system.
The quantity $N_{\text{pair}}^{r<s}$ is the number of 
pairs with interspecies 
distances smaller than $s$. For zero-range interactions, $s$
is taken to zero. For finite-range interactions, in contrast,
$s$ goes to a small value such that $s$ is larger than the range $r_0$ 
of the underlying two-body potential.
The pair relation can be written in terms of the 
short distance behavior of the pair distribution
function $P_{12}(r)$ for the spin-up---spin-down 
pairs~\cite{Tan1,Tan2,Tan3,DailyBlume}.
Throughout, we employ the normalization
$4 \pi \int_0^{\infty} P_{12}(r) r^2 dr =
N_{1} N_{2}$.
The thermally averaged expectation values are obtained by 
employing two complementary approaches, a ``microscopic approach''
and a ``direct approach''.

In the microscopic approach, the thermal expectation value of an observable
$A$
is obtained using 
$\langle A \rangle_{\text{th}} = \sum_j \exp[-E_j/(k_BT)] A_j / 
\sum_j\exp[-E_j/(k_BT)]$,
where the sum runs over all 
eigen energies $E_j$ (with associated eigen states $\psi_j$)
of the Hamiltonian $H$ and
$A_j=\langle \psi_j|A|\psi_j \rangle / \langle \psi_j| \psi_j\rangle$.
The solutions to the time-independent Schr\"odinger equation
are obtained semi-analytically for the interaction model
$V_{\text{F}}$~\cite{busch,WernerCastin} 
and using a basis set expansion approach
for the interaction model 
$V_{\text{G}}$~\cite{cgbook,DailyBlume10,rakshit12}.
The direct approach is based on calculating $\langle A \rangle$
from the density matrix $\rho$, 
$\langle A \rangle = {\text{Tr}}(A \rho) / {\text{Tr}} (\rho)$.
To sample $\rho$, we 
employ the path integral Monte Carlo (PIMC)
approach~\cite{CeperleyRev}.
Because of the Fermi sign problem~\cite{SignProblem},
the applicability of this approach is expected to be limited
to the high temperature regime.

\begin{figure}
\centering
\includegraphics[angle=0,width=0.4\textwidth]{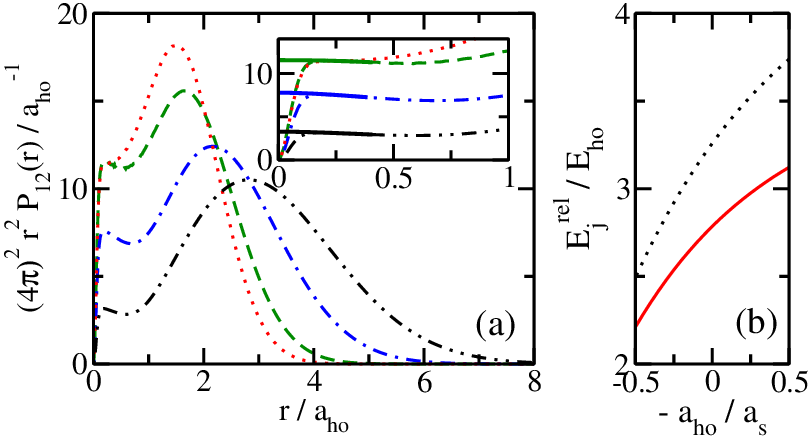}
\caption{(Color online)
(a)  
The dotted, 
 dashed,
dash-dotted
and  dash-dot-dotted
lines show the scaled pair
  distribution function 
$r^2 P_{12}(r)$
for $k_BT/(\hbar\omega) = 0,0.6,1.2$ and $2$,
  respectively, for
 the $(3,1)$ system interacting through $V_{\text{G}}$
with $r_0=0.06a_{\text{ho}}$.
The $T=0$ curve
is determined using the basis set expansion approach 
while the finite $T$ curves are determined using the
PIMC approach.
The thick solid lines in the inset
of panel~(a), which is a
blow-up of the small $r$ region, show the extrapolation to $r=0$.
(b) 
The solid and dotted lines
show the relative energy of the ground 
state with $L^{\Pi}=1^-$
and first excited state with $L^{\Pi}=0^+$ of 
the $(2,1)$ system 
interacting through $V_{\text{G}}$
with $r_0=0.06a_{\text{ho}}$ as a function of $-1/a_s$.
 }\label{FigPairDistribution31System}
\end{figure}

Figure~\ref{FigPairDistribution31System}(a)
shows 
the scaled pair distribution function 
$r^2 P_{12}(r)$  for the $(3,1)$
system for four temperatures,
$k_BT/ E_{\text{ho}}=0$, $0.6$, $1.2$ and $2$.
At $T=0$ [dotted line in Fig.~\ref{FigPairDistribution31System}(a)],
$P_{12}$ is governed by the lowest eigenstate, which has
$L^{\Pi}=1^+$ symmetry~\cite{rakshit12} ($L$ and $\Pi$ denote
the orbital angular momentum and parity, respectively).
As the temperature increases,
excited states contribute. The second lowest state has
$1^-$ symmetry. 
Compared to the ground state,
its $P_{12}$ has an increased amplitude in the small but
finite $r$ region.
The scaled pair distribution function $r^2 P_{12}$ for
$k_BT=0.6 E_{\text{ho}}$ (dashed line) has a comparable 
amplitude to that for $T=0$; however, clear differences are visible at
larger interspecies distances $r$.
For yet larger $T$, the small $r$ amplitude decreases drastically
[see dash-dotted and dash-dot-dotted lines in Fig.~\ref{FigPairDistribution31System}(a)]
while the maximum of $r^2 P_{12}$ moves to larger $r$.
To extract the contact from $r^2 P_{12}$,
we fit the small $r$ region ($r$ 
larger than $r_0$) 
and extrapolate
the fit to $r=0$ [see thick
solid lines in the inset of Fig.~\ref{FigPairDistribution31System}(a)].

\begin{figure}
\centering
\includegraphics[angle=0,width=0.35\textwidth]{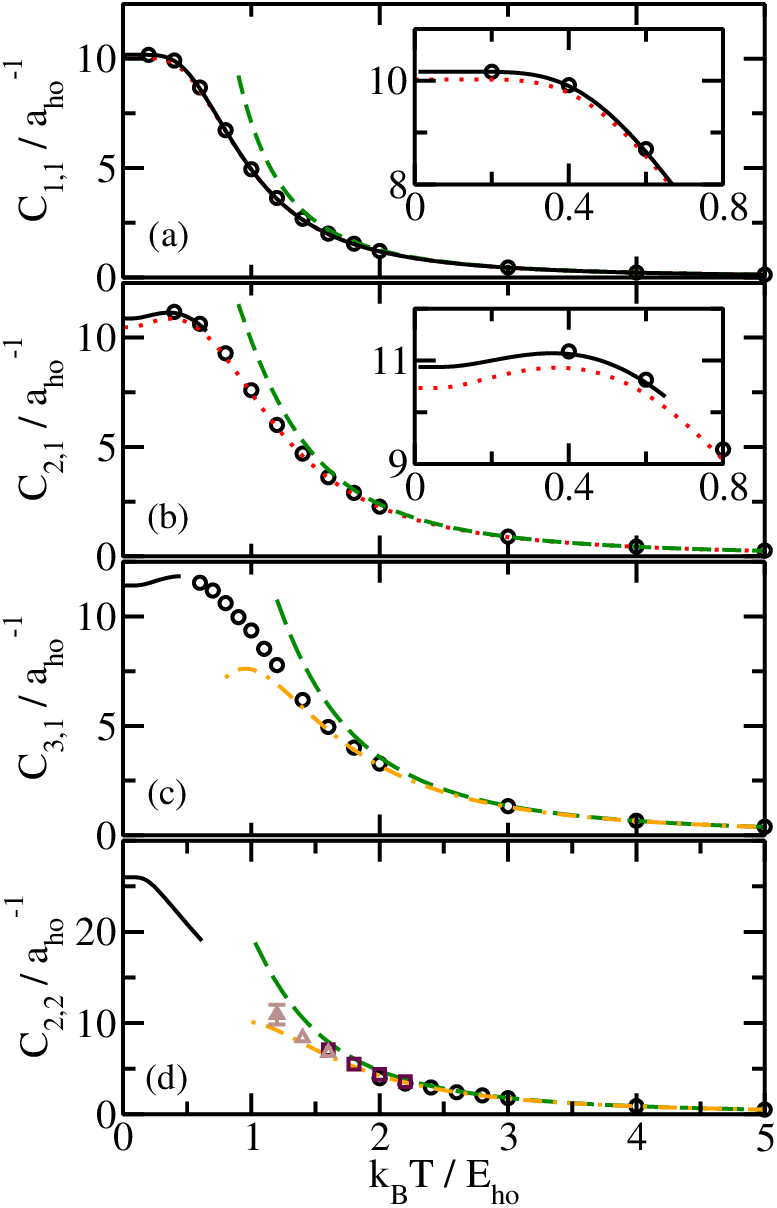}
\caption{(Color online)
  Contact $C_{N_1,N_2}$ as a function of $k_BT/E_{\text{ho}}$ 
  for the (a) $(1,1)$, (b) $(2,1)$, 
(c) $(3,1)$, and (d) $(2,2)$ systems. 
  The circles, squares and triangles show $C_{N_1,N_2}$ for $V_{\text{G}}$
  with $r_0/a_{\text{ho}}=0.06$, $0.08$ and $0.1$ obtained using the
PIMC approach.
  The solid lines show $C_{N_1,N_2}$ for $V_{\text{G}}$
with $r_0=0.06a_{\text{ho}}$
obtained using the basis set expansion approach.
  For comparison,
the dotted lines in panels (a) and (b) show
$C_{N_1,N_2}$ obtained using $V_{\text{F}}$.
  (a) The
dashed line shows the first-order Taylor expansion at high temperature.
  (b) The dashed line shows the cluster expansion, i.e.,
the quantity $2 C_{1,1}$.
  (c)/(d) The dashed and dash-dotted lines 
show 
the leading order term of the
cluster expansion and the full cluster expansion, respectively.
  The insets of panels (a) and (b) show blow-ups
of the low temperature regions.
 }\label{FigIndividualContact}
\end{figure}

Figure~\ref{FigIndividualContact} 
shows the contact
$C_{N_1,N_2}$ at unitarity for $N=2-4$
as a function of the temperature.
The symbols show the PIMC results, obtained by analyzing the 
scaled pair 
distribution functions $r^2 P_{12}(r)$ for $V_{\text{G}}$
with $r_0 \ll a_{\text{ho}}$.
The solid lines in Fig.~\ref{FigIndividualContact}
show the contact for $r_0=0.06a_{\text{ho}}$
obtained by evaluating the adiabatic relation via the microscopic approach.
It can be seen that the contact 
calculated by evaluating the adiabatic
relation via the microscopic approach and the
pair relation via the direct approach
agree or connect smoothly for the three system sizes considered.

To estimate the dependence of the contact on the range
$r_0$ of the underlying 
two-body potential,
we determine the contact of the $(1,1)$ and $(2,1)$ systems
with zero-range interactions (see supplemental material).
The dotted lines in Figs.~\ref{FigIndividualContact}(a)
and \ref{FigIndividualContact}(b) show the result.
In the low temperature regime, 
the contact for $r_0=0$ lies below that for $r_0>0$ for
the $(1,1)$ and $(2,1)$ systems.
At $k_BT = 0.4 E_{\text{ho}}$, e.g., the
$(1,1)$ and $(2,1)$ contacts for $r_0=0.06a_{\text{ho}}$
lie 
1.5\% and 3\%, 
respectively, above the contact for
$r_0=0$.
At large $T$,
the dependence of the contact on the range 
is negligibly small.
Our 
PIMC simulations 
suggest a similar range dependence for the $(3,1)$, $(4,1)$ and $(2,2)$
systems.

Figures~\ref{FigIndividualContact}(a)-\ref{FigIndividualContact}(d) 
show an intriguing 
dependence of the contact on the temperature.
$C_{1,1}$ and $C_{2,2}$
decrease monotonically with increasing temperature while
$C_{2,1}$ and $C_{3,1}$ exhibit a maximum at $k_BT \approx 0.36 E_{\text{ho}}$
and between $0.4 E_{\text{ho}}$ and $0.5 E_{\text{ho}}$, respectively.
To explain this behavior, it is instructive to evaluate
the adiabatic relation via the microscopic approach.

For the $(1,1)$ system with zero-range interactions,
one finds $\partial
E_j/(\partial(-a_s^{-1}))=\Gamma(j+1/2)2^{3/2}/[\pi(2j)!] E_{\text{ho}} a_{\text{ho}}$
 for the $s$-wave states and
$\partial E_j/(\partial(-a_s^{-1}))=0$ for all higher partial wave
states~\cite{busch,blume12}.
The fact that $C_{1,1}$ decreases monotonically with decreasing temperature
is thus a direct consequence of the fact that 
$\partial E_j/(\partial(-a_s^{-1}))$ (for $s$-wave
states) decreases with increasing $j$.
The inclusion of effective range corrections does not,
if applied to the Gaussian model interaction with sufficiently small
$r_0$, change this picture~\cite{supplement}.
A similar analysis, based on the numerically determined energies,
holds for the $(2,2)$ system.
Figure~\ref{FigPairDistribution31System}(b)
shows the lowest two relative eigen energies, which
correspond to states  
with $L^{\Pi}=1^-$ and $0^+$ symmetry, respectively, of the
$(2,1)$ system as a function of 
$-a_{\text{ho}}/a_s$ for $r_0=0.06a_{\text{ho}}$.
The slope of the $1^-$ state is smaller than that
of the $0^+$ state.
This can be understood as follows.
In the $a_s \rightarrow 0^-$ limit, the lowest state has $L^{\Pi}=1^-$
symmetry.
In the $a_s \rightarrow 0^+$ limit, in contrast,
the lowest state has $L^{\Pi}=0^+$
symmetry.
The two states cross at 
$a_{\text{ho}}/a_s \approx 1$ ~\cite{kestner,kolck,stecher}.
Correspondingly, in the unitary regime the energy of the lowest
$L=0$ state changes more rapidly
with $-1/a_s$ than that of the lowest $L=1$ state,
implying that the contact 
at unitarity increases in the low temperature regime
where the inclusion of only two states yields converged results.
A more comprehensive analysis that accounts for all states is
presented in the supplemental material~\cite{supplement}.
For the $(3,1)$ system, a similar argument can be made in the low temperature
regime where the inclusion of just a few states suffices.
The calculations presented here suggest that $C_{N_1,N_2}$
decreases monotonically 
with $T$ if $N_1-N_2=0$ and exhibits a maximum at finite $T$
if
$N_1-N_2 \ne 0$. 
While it is tempting to generalize these results to larger systems,
it should be noted that the density of states increases dramatically
with increasing $N$ and that application of a few-state model
will be limited to smaller temperatures as $N$
increases.

We now introduce a 
high temperature cluster expansion of the contact at unitarity.
A formal discussion of
the cluster expansion in the canonical ensemble applied to classical systems
is provided in
Ref.~\cite{mathpaper}.
The $(N_1,N_2)$ system contains $N_1 N_2$ 
interacting pairs and one might expect
that, using 
the argument
that two-component Fermi gases behave 
universally~\cite{giorginiReview,blumerev},
the high temperature tail of $C_{N_1,N_2}$ is governed 
by $N_1 N_2 C_{1,1}$
for $k_BT \gg E_{\text{ho}}$
[see dashed lines in 
Figs.~\ref{FigIndividualContact}(b)-\ref{FigIndividualContact}(d)].
The next term in the cluster expansion,
applicable to systems with $N>3$, 
depends on the ``three-body term''
$C_{2,1}-2C_{1,1}$,
\begin{eqnarray}
\label{eq_clusterexpansion}
\frac{C_{N_1,N_2}}{N_1 N_2}= C_{1,1} +
\frac{N_1+N_2-2}{2}(C_{2,1}-2 C_{1,1})+\cdots.
\end{eqnarray}
The dashed and dash-dotted lines in 
Figs.~\ref{FigIndividualContact}(c) and \ref{FigIndividualContact}(d)
show the leading term and the sum of the leading and sub-leading terms
for the $(3,1)$ and $(2,2)$ systems.
The inclusion of the three-body term improves the validity 
regime of the cluster expansion notably. 
Assuming zero-range interactions,
the leading order terms of the
Taylor expansions of $C_{1,1}$ and $C_{2,1}-2C_{1,1}$
around $\tilde{\omega} \ll 1$,
where $\tilde{\omega}=E_{\text{ho}}/(k_BT)$, 
are $4\pi^{1/2}\tilde{\omega}^{5/2} a_{\text{ho}}^{-1}$ and
$-7.012(12)\tilde{\omega}^{11/2} a_{\text{ho}}^{-1}$, 
indicating that the three-body term is 
suppressed by $\tilde{\omega}^3$ compared to the leading order two-body term.
Figures~\ref{FigIndividualContact}(c) and \ref{FigIndividualContact}(d) 
show
that the numerically obtained $C_{3,1}$ and $C_{2,2}$ contacts
(symbols) lie above the cluster prediction (dash-dotted line),
suggesting that the corresponding leading order 
four-body expansion coefficients
are positive.
The above expansions can be viewed as canonical analogs of the virial equation
of state description of the contact within the grand canonical 
ensemble~\cite{liurev,liuNJP,supplement}. 

\begin{figure}
\centering
\includegraphics[angle=0,width=0.35\textwidth]{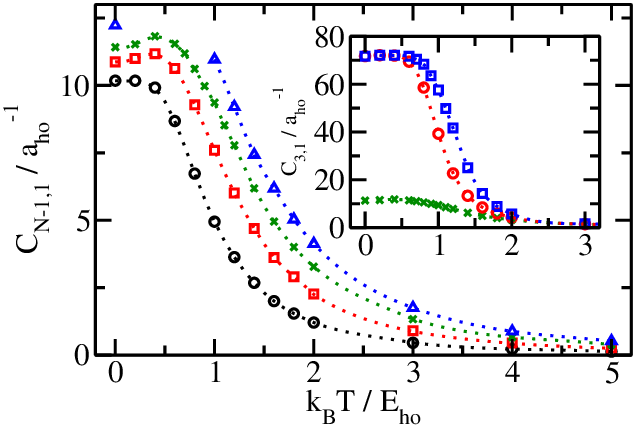}
\vspace{0.2in}
\caption{(Color online)
  Circles, squares, crosses and triangles show 
$C_{N_1,N_2}$ for the $(1,1)$, $(2,1)$,
  $(3,1)$ and $(4,1)$ systems, respectively,
interacting through $V_{\text{G}}$ with
$r_0=0.06a_{\text{ho}}$ as a function of 
$T$~\protect\cite{footnote41system}.
  Dotted lines serve as a guide to the eye.
Inset:  Crosses, squares and circles show the contact of the
$(3,1)$ system for Fermi, Boltzmann and Bose statistics,
  respectively.
}\label{FigCombinedContact}
\end{figure}

Equation~(\ref{eq_clusterexpansion}) shows that the 
contact $C_{N-1,1}$
with $N>2$ is $N-1$ times larger than $C_{1,1}$ in the 
high temperature limit.
In the low temperature limit (see Fig.~\ref{FigCombinedContact}),
in contrast, $C_{N-1,1}$ with $N>2$ is only slightly larger
than $C_{1,1}$, reflecting the fact that, to leading order,
the system can form one and not $N-1$ bound pairs.
It is also interesting to compare the limiting behaviors
of $C_{3,1}$ and $C_{2,2}$. 
$C_{2,2}$ is $4/3$ times larger 
than $C_{3,1}$ at large $T$ [see Eq.~(\ref{eq_clusterexpansion})] 
but 
roughly two times larger at low $T$.
The latter reflects the fact that the $(2,2)$ and $(3,1)$ systems can form
two dimers and one dimer, respectively.

To elucidate the role of the Fermi statistics, we 
focus on the $(3,1)$ system at unitarity with $r_0=0.06 a_{\text{ho}}$. 
The inset of Fig.~\ref{FigCombinedContact} shows
the contact obtained by treating the majority 
particles as identical fermions (crosses; these are the same 
data as discussed above),
as identical bosons (circles) and as distinguishable ``Boltzmann
particles'' (squares).
In the high
temperature regime, the
Fermi and Bose statistics can be treated as a  
correction to the Boltzmann statics. 
In the low temperature regime, in contrast, appreciable 
differences are revealed. 
The $(3,1)$ systems with Bose and Boltzmann statistics 
share the same ground state and thus
the same contact in the zero temperature limit.
Compared to the contact of the Bose and Boltzmann systems, 
that of the Fermi system is strongly
suppressed as a consequence of the Pauli exclusion principle. 
Specifically, the $(3,1)$ Fermi system at unitarity does not support a 
bound state in free space while the $(3,1)$ Bose and Boltzmann
systems do. The existence of self-bound states leads to an increased
amplitude of the pair distribution function at small interspecies distances.
Moreover, the 
Bose and Boltzmann systems are---unlike the Fermi system---not 
fully universal, i.e., their properties are, in addition
to the $s$-wave scattering length, governed by
a three-body parameter. This implies that the Bose and Boltzmann
systems are characterized by a non-zero three-body contact
in addition to the two-body contact considered throughout this 
paper~\cite{BraatenPlatter,footnote1}.

Finite-temperature effects
play an important role in many finite-sized systems,
including
atomic clusters, nuclei and quantum dots.
Our work demonstrates that small harmonically trapped two-component
Fermi gases
with infinitely strong interspecies
$s$-wave interactions,
which can be realized and probed experimentally,
also exhibit intriguing 
dependencies on the temperature.
In particular, we proposed a high-temperature cluster expansion
in the canonical ensemble,
quantified the range dependence of the contact, observed and 
interpreted the distinctly different behavior of the
contact of spin-balanced and spin-imbalanced Fermi gases in the low temperature
regime, and elucidated the role of the Fermi statistics.
Throughout, we reported the temperature in terms of the
natural energy scale of the harmonic oscillator. 
Other relevant temperature scales
are the Fermi temperature $T_F$
and the critical temperature
$T_c$, $T_F=2.5 E_{\text{ho}}/k_B$ for $N=3-5$~\cite{footnotetf}  and
$T_c\approx 0.2T_F$ for the trapped spin-balanced system~\cite{Stringari13}.
Our calculations cover temperatures much smaller and much larger
than these characteristic temperature scales.
Future studies will be aimed at determining the critical temperature, 
and the superfluid fraction
and superfluid density of small trapped Fermi gases.

Support by the National
Science Foundation (NSF) through Grant No.
PHY-1205443
is gratefully acknowledged.
This work used the Extreme Science and Engineering
Discovery Environment (XSEDE), which is supported by
NSF grant number OCI-1053575, and the
WSU HPC.
This work was additionally supported by the 
NSF through a grant for the Institute 
for Theoretical Atomic, Molecular and Optical Physics 
at Harvard University and
Smithsonian Astrophysical Observatory.

\onecolumngrid


\begin{thebibliography}{10}
\bibitem{footnote41system}
For the $(4,1)$ system, the results
derived from the basis set expansion approach cover the temperature range
$k_BT \lesssim0.1 E_{\text{ho}}$.
The results derived from the PIMC approach cover the temperature range
$k_BT \gtrsim  E_{\text{ho}}$.
For this system, we 
did not treat the temperature
regime between $0.1 E_{\text{ho}} \lesssim k_BT \lesssim E_{\text{ho}}$.
Nevertheless, our results suggest that $C_{4,1}$ exhibits a maximum at finite
$T$, like the other maximally polarized systems considered in
this paper.
\bibitem{footnote1}
The three-body contact~\cite{BraatenPlatter} 
should not be confused with the two-body contact
$C_{N_1,N_2}$ of the two-, three- and higher-body systems.
For the Fermi systems considered in this work, 
the three-body contact vanishes 
identically.
\bibitem{footnotetf}
The Fermi temperature
of small harmonically trapped
two-component Fermi gases
is defined through the energy of the highest single-particle state
of the non-interacting system.
\end{thebibliography}
\end{document}


\title{Supplemental material for ``Harmonically trapped Fermi gas: Temperature dependence of the
Tan contact''}
\author{Yangqian Yan}
\affiliation{Department of Physics and Astronomy,
Washington State University,
  Pullman, Washington 99164-2814, USA}
\author{D. Blume}
\affiliation{Department of Physics and Astronomy,
Washington State University,
  Pullman, Washington 99164-2814, USA}
\affiliation{ITAMP, Harvard-Smithsonian Center for Astrophysics,
Cambridge, Massachusetts 02138, USA}
\date{\today}

\maketitle

The notation used in this supplemental material follows that
introduced in the main text.

\subsection{(1,1) system with zero-range interactions}
The relative energies of the $(1,1)$ system with $s$-wave zero-range
interactions can be determined by solving the transcendental 
equation~\cite{busch}
\begin{eqnarray}
\label{eq_busch}
\frac{\sqrt{2}\Gamma(3/4-E^{\text{rel}}/(2
E_{\text{ho}}
))}
{\Gamma(1/4-E^{\text{rel}}/(2
E_{\text{ho}}
))} 
= \frac{a_{\text{ho}}}{a_s}.
\end{eqnarray}
At unitarity, the relative $s$-wave energies
read $E^{\text{rel}}_j = (2j + 1/2) 
E_{\text{ho}}$, where $j=0,1,\cdots$.
States with relative orbital angular momentum
$L$ greater than zero are not affected by the interactions.
Using these energies and the 
expressions for the change of the relative energies with
$-1/a_s$ from the main text, one finds
\begin{equation}
C_{1,1}(\tilde{\omega})
=  8\sqrt{\pi}\frac{ e^{\tilde{\omega }} \left(e^{\tilde{\omega }}-1\right)^2
  \sqrt{e^{-\tilde{\omega }} \sinh \left(\tilde{\omega }\right)}}{
  e^{\tilde{\omega }} \left[e^{\tilde{\omega }} \left(e^{\tilde{\omega
  }}-2\right)+4\right]-1} a_{\text{ho}}^{-1}
  \label{Eq11Contact}
\end{equation}
[see dotted line in Fig.~2(a) of the main text].

To quantify the corrections that arise from the 
finite range of the interspecies interaction potential, 
we consider three approaches:

{\em{(i)}}
Using a B-spline approach, we calculate the relative energies
of the $(1,1)$ system as a function of $-1/a_s$ up to $500 E_{\text{ho}}$ for
the first 50 angular momentum channels.
Using these energies and the corresponding slopes, we
calculate the contact numerically. 
The solid line in Fig.~2(a) of the main text
shows the result for the Gaussian potential with $r_0=0.06 a_{\text{ho}}$.

{\em{(ii)}} 
We replace $-1/a_s$ in Eq.~(\ref{eq_busch})
by $-1/a_s + r_{\text{eff}} k^2 /2$,
where $r_{\text{eff}}$ denotes the effective range and $k$
is related to the relative energy via $\hbar^2 k^2/m=E^{\text{rel}}$.
The leading order effective range
corrections 
to the energy
and to the change of 
the energy at unitarity can then be derived analytically~\cite{blume12}.
Using these expressions, the contact can be readily
\begin{figure}
\includegraphics[angle=0,width=0.35\textwidth]{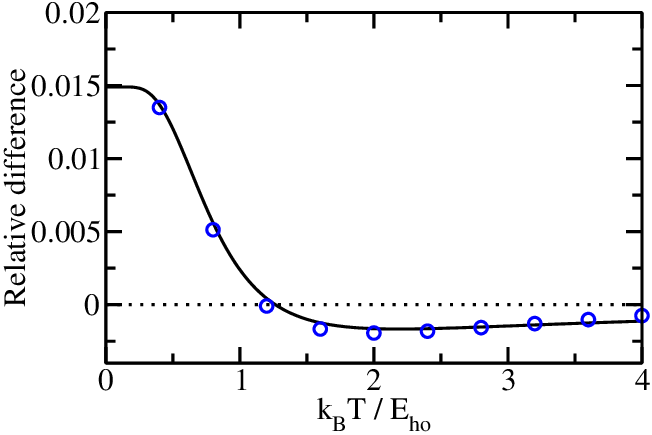}
\centering
\vspace{0.2in}
\caption{(Color online)
Range dependence of the
contact $C_{1,1}$ 
for $r_0=0.06a_{\text{ho}}$
as a function of $k_B T / E_{\text{ho}}$.
The circles 
show the difference between the 
finite-range contact 
calculated
using approach~{\em{(i)}} 
and the
zero-range contact,
normalized by the zero-range contact.
The solid line 
shows the difference between the 
finite-range contact 
calculated
using approach~{\em{(ii)}}
and the
zero-range contact,
normalized by the zero-range contact.
In the large $T$ limit, the difference approaches zero from below.
In approach {\em{(ii)}},
$r_{\text{eff}}=0.12178 a_{\text{ho}}$---corresponding to $V_{\text{G}}$ with 
$r_0=0.06 a_{\text{ho}}$---has been used.
 }\label{fig_supplement1}
\end{figure} 
determined numerically within the microscopic approach.
On the scale of Fig.~2(a) of the main text, the resulting contact
would be indistinguishable from the solid line.

{\em{(iii)}}
To account for the fact that higher partial wave channels are
affected by the interspecies interactions if the range of the underlying
two-body potential is finite, we generalize the above effective range
treatment to finite angular momenta. Specifically, we replace the 
generalized scattering lengths in the transcendental equations for the higher
partial waves~\cite{deutsch} 
by the corresponding effective range expansions
and determine the corrections to the energy and to the
slope of the energy
for vanishing generalized scattering lengths due to the
effective range.
Calculating the effective ranges for the Gaussian potential
with $r_0=0.06a_{\text{ho}}$ for the lowest
few partial wave channels, 
we find that 
the effective range correction of the contact
due to the $L=1$ channel 
is about 0.0003\% and 0.015\% 
for $k_BT/E_{\text{ho}}=1/2$ and $2$, respectively.
For fixed $T$,
the corrections decrease with increasing $L$.
Although the $L=1$ channel leads to a larger percentage correction
at large $T$ than at small $T$,
its overall role is negligible since the contact itself is small in the large
$T$ regime.

Overall, the agreement between the contacts calculated using approaches 
{\em{(i)}}-{\em{(iii)}} is excellent.
To bring out the size of the finite-range corrections,
Fig.~\ref{fig_supplement1} shows the
difference between the finite-range and zero-range
contacts, normalized by the zero-range contact.
The circles show the relative difference for the case where the
finite-range contact is calculated using approach {\em{(i)}}
while the solid line shows the
relative difference for the case where the
finite-range contact is calculated using approach {\em{(ii)}}.
The difference is largest at low $T$ and decreases rapidly.
The difference changes sign at $k_BT \approx 1.25 E_{\text{ho}}$
and then approaches zero from the negative side.
For $k_BT \gtrsim  E_{\text{ho}}$, 
the percentage deviation is smaller than 0.5\% and thus,
for all practical purposes, negligible.

A key result of the above analysis is that the two-body
contact is nearly fully determined by the $s$-wave channel and that higher
partial wave contributions play a negligible role
if the interspecies interactions are modeled by the Gaussian 
potential $V_{\text{G}}$ with $r_0=0.06 a_{\text{ho}}$.
This is crucial for our analysis of the $(2,1)$ system
discussed in the next section, which assumes interspecies 
$s$-wave zero-range interactions.
It also suggests that our high temperature results for larger systems,
obtained by using $V_{\text{G}}$ with $r_0=0.06 a_{\text{ho}}$,
are very close to the universal zero-range results.

\subsection{(2,1) system with zero-range interactions}
To determine the contact of the $(2,1)$
system with zero-range interactions
at
unitarity, we resort to the
hyperspherical coordinate approach~\cite{WernerCastin,seth10}.
In this approach, the solutions of the relative
Schr\"odinger equation are obtained in a two step
process. First, the hyperangular
Schr\"odinger equation is solved for fixed hyperradius $R$.
Second, a set of 
coupled hyperradial
Schr\"odinger equations is solved.
For zero-range interactions with infinitely
large $s$-wave scattering length,
the coupling between the hyperangular and hyperradial
degrees of freedom vanishes and the relative
eigen energies at unitarity
can be written as 
$E^{\text{rel}}=(2q+s_{L,\nu}+1)
E_{\text{ho}}
$,
where the hyperradial quantum number $q$ takes the values
$0,1,\cdots$.
The $s_{L,\nu}$
are solutions of the transcendental equation~\cite{WernerCastin,seth10}
\begin{eqnarray}
  \frac{ (-1)^L \, _2F_1\left(\frac{1}{2} (L-s_{L,\nu}+2),\frac{1}{2}
  (L+s_{L,\nu}+2);L+\frac{3}{2};\frac{1}{4}\right)}{\pi  (2 L+1)\text{!!}} + \nonumber \\
\frac{1}{\Gamma
  \left(\frac{1}{2} (L-s_{L,\nu}+1)\right) \Gamma \left(\frac{1}{2}
(L+s_{L,\nu}+1)\right)} \nonumber \\
= \frac{1}{\sqrt{2}\Gamma \left(\frac{1}{2} (L-s_{L,\nu}+2)\right) \Gamma \left(\frac{1}{2} 
(L+s_{L,\nu}+2)\right)}\frac{R}{a_s}
\label{eq_svalue}
\end{eqnarray}
for $1/a_s=0$.
In Eq.~(\ref{eq_svalue}),
the hyperradius $R$ 
is defined through
$R^2=2(r_{12}^2+r_{23}^2+r_{13}^2)/3$ and
$_2F_1$ denotes the hypergeometric function.

To determine the change of the relative energies 
at unitarity with $-1/a_s$,
we 
replace $s_{L,\nu}$ in Eq.~(\ref{eq_svalue}) by
$s_{L,\nu} + \Delta s_{L,\nu}$, where
$|\Delta s_{L,\nu}|$ is assumed to be small. 
Taylor expanding Eq.~(\ref{eq_svalue})
around the known $s_{L,\nu}$ value,
we find 
$\Delta s_{L,\nu}= c_{L,\nu} R/a_s$.
We insert this into the effective
hyperradial potential
$\hbar^2 [(s_{L,\nu}+\Delta s_{L,\nu})^2-1/4]/(2mR^2)$
and treat the leading order correction as a perturbation,
i.e., we define $V_{\text{pert}}(R)=\hbar^2 c_{L,\nu} s_{L,\nu} /(mR a_s)$.
For each $s_{L,\nu}$, we calculate the exact proportionality
constant
$c_{L,\nu}$
and determine the change of the relative energy 
by treating $V_{\text{pert}}(R)$  in
first-order perturbation theory in the hyperradial
Schr\"odinger equation, i.e., by evaluating
the integral
$\int_0^\infty |F_{q, s_{L, \nu}}(R)|^2 V_{\text{pert}}(R) dR$.
The unperturbed radial wave functions
$F_{q, s_{L,\nu}}(R)$ 
are obtained by solving the
hyperradial Schr\"odinger equation for the unitary problem.
Using the known expressions for $F_{q, s_{L,\nu}}(R)$~\cite{WernerCastin},
the integral can be evaluated analytically,
yielding the leading-order variation of $E^{\text{rel}}$
with $-1/a_s$ for each $s_{L,\nu}$ and $q=0,1,\cdots$.
The approach outlined here reproduces the recurrence
relations of Refs.~\cite{castin,Moroz12}.
Using Mathematica, the energies and slopes of the energies
of the $(2,1)$
system 
at unitarity
can be calculated with essentially
arbitrary accuracy, thereby allowing us to calculate
the temperature-dependent contact with high accuracy.
In calculating the partition function,
care has to be exercised as the above formalism excludes a large number
of ``trivial energy levels'' that are not affected by the $s$-wave 
interactions~\cite{WernerCastin}. We account for these ``trivial states''
using the ideas discussed in Ref.~\cite{DailyBlume10}.

Table~\ref{tab_supplement1}
tabulates the contact $C_{2,1}$ [see also dotted line
in Fig.~2(b) of the main text].
\begin{table}
\caption{Contact $C_{2,1}$ as a function of temperature for $s$-wave zero-range
interactions at unitarity.
}
\label{tab_supplement1}
\centering
\begin{tabular}{c|c}
\hline
\hline
 $k_B T / E_{\text{ho}}$  & $C_{2,1}/ a_{\text{ho}}^{-1}$ \\
\hline
0.1 & 10.4986\\

0.2 & 10.6717\\

0.3 & 10.8225\\

0.4 & 10.8529\\

0.5 & 10.7047\\

0.6 & 10.3479\\

0.7 & 9.79272\\

0.8 & 9.08421\\

0.9 & 8.28504\\

1.0 & 7.45680\\

1.1 & 6.64831\\

1.2 & 5.89176\\

1.3 & 5.20435\\

1.4 & 4.59212\\

1.5 & 4.05395\\

1.6 & 3.58469\\

1.7 & 3.17735\\

1.8 & 2.82444\\

1.9 & 2.51872\\

2.0 & 2.25359\\

2.1 & 2.02320\\

2.2 & 1.82248\\

2.3 & 1.64710\\

2.4 & 1.49338\\

2.5 & 1.35820\\

2.6 & 1.23893\\

2.7 & 1.13335\\

2.8 & 1.03959\\

2.9 & 0.956067\\

3.0 & 0.881422\\

3.1 & 0.814516\\

3.2 & 0.754375\\

3.3 & 0.700163\\

3.4 & 0.651166\\

3.5 & 0.606768\\

3.6 & 0.566439\\

3.7 & 0.529719\\

3.8 & 0.496209\\

3.9 & 0.465562\\

4.0 & 0.437476\\

4.1 & 0.411684\\

4.2 & 0.387954\\

4.3 & 0.366081\\

4.4 & 0.345884\\

4.5 & 0.327202\\

4.6 & 0.309894\\

4.7 & 0.293834\\

4.8 & 0.278908\\

4.9 & 0.265017\\

5.0 & 0.252072\\
\hline
\hline
\end{tabular}
\end{table} 
These zero-range
results serve as a benchmark for our PIMC simulations.
Moreover, the semi-analytic expressions for
the energy and its variation with $-1/a_s$
can also be used to calculate
the
third-order contact coefficient~\cite{liuNJP}.
We find
$0.0269223(3)$,
which notably improves upon the 
accuracy of the value of $0.02692(2)$ of Ref.~\cite{liuNJP};
a more detailed discussion of the connection
between the second- and third-order
contact coefficients~\cite{liuNJP} and our cluster expansion
is given below.

\begin{figure}
\centering
\includegraphics[angle=0,width=0.4\textwidth]{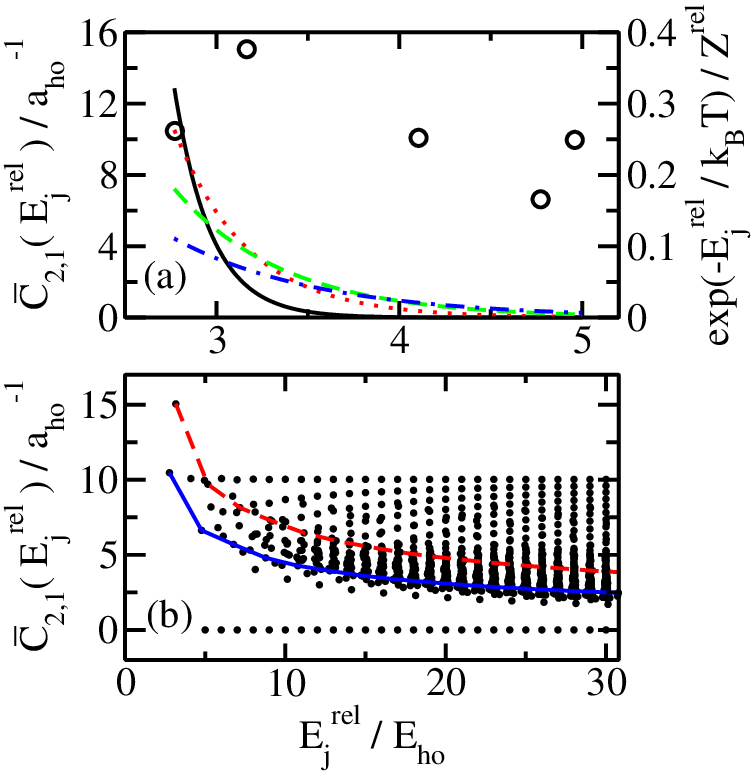}
\vspace{0.2in}
\caption{(Color online)
(a) The solid, dotted, dashed and dash-dotted lines 
show the weight factor 
$\exp[-E_j^{\text{rel}}/ (k_BT)]/Z^{\text{rel}}$
(see right axis) as a
function of the relative energy 
for $k_BT/E_{\text{ho}}=0.2, 0.4, 0.6$ and $0.8$, respectively.
The circles show the zero-temperature contact 
$\bar{C}_{2,1}$ (see left axis) for states with
$E_j^{\text{rel}}< 5 E_{\text{ho}}$. (b)
The dots show the zero-temperature contact 
$\bar{C}_{2,1}$ as a function of the
relative energy for $E_j^{\text{rel}}
\lesssim
30
E_{\text{ho}}$.
The solid (dashed) line connects 
the contacts of $L=1$ ($L=0$) states, 
which are characterized by the same $s_{L,\nu}$ value
but different
$q$ values; the $s_{L,\nu}$ values chosen are the smallest ones for
both $L=1$ and $0$.
 }\label{fig_supplement2}
\end{figure}

We now refine the two-state model, which was used in the
main text to explain why the maximum of $C_{2,1}$ occurs at finite $T$.
Circles in Fig.~\ref{fig_supplement2}
show the zero-temperature contacts $\bar{C}_{2,1}(E_j^{\text{rel}})$,
$\bar{C}_{2,1}(E_j^{\text{rel}})  
= (4 \pi m / \hbar^2) \partial E^{\text{rel}}_j / (\partial(-a_s^{-1}))$,
associated with the $j$th eigenstate
of the $(2,1)$ system at unitarity as a function of $E^{\text{rel}}$.
Panel~(a) focuses on the small $E^{\text{rel}}$ region
while panel~(b) extends up to $E^{\text{rel}}=30 E_{\text{ho}}$.
Figure~\ref{fig_supplement2} shows that the 
zero-temperature contacts are non-negative.
Moreover, all $\bar{C}_{2,1}(E_j^{\text{rel}})$ but one 
are smaller than $11a_{\text{ho}}^{-1}$. The ``outlier'' corresponds to the 
lowest $L=0$ state, i.e.,  
the state with the second lowest eigen energy.
To calculate the temperature-dependent contact
$C_{2,1}$  within the microscopic
approach, the zero-temperature contacts $\bar{C}_{2,1}(E_j^{\text{rel}})$
need to be weighted by the temperature-dependent
Boltzmann factors $\exp[-E_j^{\text{rel}}/(k_B T)]/Z^{\text{rel}}(T)$,
where $Z^{\text{rel}}(T)$ denotes the partition
function that accounts for  the relative degrees of freedom.
Lines
in Fig.~\ref{fig_supplement2}(a)
show these ``weight factors'' 
for four different $T$, $k_B T/E_{\text{ho}}=0.2,0.4,0.6$ and $0.8$.
It can be seen that the weight factor drops off quickly for 
$k_BT/E_{\text{ho}}=0.2$, indicating that the ``outlier''
as well as the zero-temperature contacts of the other excited
states
contribute negligibly to $C_{2,1}$.
At $k_BT/E_{\text{ho}}=0.4$, in contrast, the ``outlier'' carries
appreciable weight compared to the zero-temperature contact of the lowest 
eigen state. At yet higher $T$, the weight factor falls off slower, 
thereby reducing the relative importance of the ``outlier''.

\subsection{Basis set expansion approach}
The explicitly correlated Gaussian basis set
expansion
approach with semi-stochastic parameter optimization has been used extensively
in the literature to describe the behavior of small 
harmonically trapped two-component Fermi gases with short-range
interactions (see, e.g., Ref.~\cite{rakshit12} for details).
The energies of the three- and four-body systems can be calculated
with an accuracy of better than 1\% (and often better than $0.1$\%).
To determine the solid lines in 
Figs.~2(b)-2(d) of the main text,
we used energy cutoffs $E^{\text{rel}}$ around $11 E_{\text{ho}}$, 
$9 E_{\text{ho}}$, 
and
$9.5 E_{\text{ho}}$, respectively.
Including the degeneracies arising from the projection
quantum number as well as the relative energies of states that are only very
weakly affected by the short-range interactions,
this amounts to around
$1250$, $230$ and $700$ energy levels
for the $(2,1)$, $(3,1)$ and $(2,2)$ systems, respectively.
The use of a finite energy cutoff implies that the determination of the 
finite-range contact within the microscopic approach is
limited to the low temperature regime.
Convergence of the contact with respect to the energy cutoff was tested 
by including successively fewer energy levels.

\subsection{Path integral Monte Carlo approach}
Our PIMC implementation largely follows that described
in Refs.~\cite{CeperleyRev,boninsegni}.
In the low temperature regime,
we find that we have to adjust the simulation parameters
that control the sampling of the unpermuted
configurations as well as those that control
the sampling of the permutations carefully.
We employ 
the second- and fourth-order Trotter formula~\cite{chin97,voth01};
higher-order propagation schemes
did not seem to lead to further improvements.

It is well known that the standard PIMC algorithm, as employed here,
suffers from the Fermi sign problem~\cite{SignProblem}.
Simply put, the signal to noise ratio decreases exponentially
with decreasing temperature and increasing number of particles.
As demonstrated by our simulation results, the Fermi sign problem is
sufficiently small for the parameter region considered in this
Letter.
In particular, the PIMC algorithm allows us to investigate a 
region of the physical parameter space (strong 
short-range interactions and relatively low temperature) that is 
inaccessible by other numerical approaches.

\subsection{Cluster expansion in canonical ensemble
and connection to virial equation of state in grandcanonical ensemble}
This section elaborates on the 
cluster expansion introduced in the main part of the text.
To further explore the high temperature behavior, 
we Taylor expand $C_{1,1}$ and $C_{2,1}-2C_{1,1}$ 
around $\tilde{\omega} \ll 1$.
We find
\begin{eqnarray}
C_{1,1} = \tilde{\omega}^{5/2}
\sum_{i=0}^{\infty} c_{1,1}^{(i)} \tilde{\omega}^{i} a_{\text{ho}}^{-1}
\label{eq_c11_canonical}
\end{eqnarray}
and 
\begin{eqnarray}
C_{2,1}-2C_{1,1} = \tilde{\omega}^{5/2}
\sum_{i=3}^{\infty} c_{2,1}^{(i)} \tilde{\omega}^{i} a_{\text{ho}}^{-1}.
\label{eq_c21_canonical}
\end{eqnarray}
Table~\ref{table_taylor}
\begin{table}
\caption{
High-temperature expansion coefficients $c_{1,1}^{(i)}$ and $c_{2,1}^{(i)}$
for the cluster expansion in the canonical ensemble.
}
\label{table_taylor}
\centering
\begin{tabular}{c|ccccc}
\hline
\hline
 $i$  &  0  & 1  
 &  2 & 3 & 4 \\
 \hline
 $c_{1,1}^{(i)}$  & $4\sqrt{\pi}$ & 0 &
 $\sqrt{\pi}/6$ & $-2\sqrt{\pi}$ & $\sqrt{\pi}/96$\\
 $c_{2,1}^{(i)}$  &  &  &  & $-7.012(12)$ & $0.0(1)$\\
\hline
\hline
\end{tabular}
\end{table} 
 lists the dimensionless coefficients
$c_{1,1}^{(i)}$ and $c_{2,1}^{(i)}$ for $i \le 4$. 

We now connect the high temperature cluster expansion 
in the canonical ensemble  to the 
more frequently used high-temperature
virial expansion in the grand canonical
ensemble~\cite{huangText}, which assumes
large number of particles.
For simplicity, we consider
 spin-balanced harmonically trapped systems, i.e.,
systems with $N_1=N_2=N/2$.
The contact $C_{\text{gc}}$ in the grand canonical virial
expansion reads~\cite{liuNJP}
\begin{equation}
  C_{\text{gc}} = 
2^{7/2}\pi^{3/2}\tilde{\omega}^{-1/2}Z_{1}(\tilde{\omega})
\left[ c_{2}(\tilde{\omega}) z^{2} + c_{3}(\tilde{\omega}) z^{3}+\cdots \right]
a_{\text{ho}}^{-1},
  \label{grandcanonical}
\end{equation}
where $Z_1(\tilde{\omega})$ denotes the partition function 
of a single particle in a spherically symmetric harmonic trap and $z$ the
fugacity, $z=\exp(\mu/k_BT)$; here, $\mu$ is the chemical potential.
The contact coefficients 
$c_{2}(\tilde{\omega})$ and $c_{3}(\tilde{\omega})$ 
for the trapped system
depend on the temperature and can be derived from the 
second- and third-order virial coefficients of the trapped system~\cite{liuNJP}.
In the following, we use the high temperature 
expansions 
\begin{eqnarray}
c_{2}(\tilde{\omega})
=\frac{1}{2\sqrt{2}\pi} \left( 1-\frac{\tilde{\omega}^2}{12} + \cdots \right)
\label{eq_contactcoeff2}
\end{eqnarray}
and 
\begin{eqnarray}
c_{3}(\tilde{\omega}) = 0.0269223(3) + \cdots .
\label{eq_contactcoeff3}
\end{eqnarray}
The high-temperature expansion of the
partition function reads
\begin{equation}
Z_1(\tilde{\omega})=  
\tilde{\omega}^{-3}\left(1-\frac{\tilde{\omega }^2}{8}+\cdots \right)
.
  \label{hightZ}
\end{equation}
The 
fugacity 
$z$ can be determined from the number equation~\cite{liuNJP}.
Expanding the number equation for small $z$, we find
\begin{equation}
  z= \frac{N}{2} \left(\tilde{\omega }^3+\frac{\tilde{\omega }^5}{8}-\frac{3}{8}  \frac{N}{2} \tilde{\omega }^6+ \cdots \right).
  \label{fugacityhight}
\end{equation}
Inserting the expansions given in 
Eqs.~(\ref{eq_contactcoeff2})-(\ref{fugacityhight})
into Eq.~(\ref{grandcanonical}),
we find
\begin{eqnarray}
  C_{\text{gc}}= 
\bigg( 
4\sqrt{\pi} \frac{N^2}{4} \tilde{\omega}^{5/2} 
+\frac{\sqrt{\pi}}{6}\frac{N^2}{4}\tilde{\omega}^{9/2}
-3\sqrt{\pi} \frac{N^3}{8}
  \tilde{\omega}^{11/2} \nonumber \\
  -1.69606(2) \frac{N^3}{8}\tilde{\omega}^{11/2}
  + \cdots
\bigg)
a_{\text{ho}}^{-1}.
  \label{grandcanonicalhight}
\end{eqnarray}
To relate Eq.~(\ref{grandcanonicalhight})
to Eqs.~(\ref{eq_c11_canonical}) and (\ref{eq_c21_canonical}),
we write, in analogy with Eq.~(4) of the main text,
\begin{eqnarray}
\frac{C_{\text{gc}}}{N^2/4}= {C}_{\text{gc};1,1}+ \frac{N-2}{2}
({C}_{\text{gc};2,1}-2 {C}_{\text{gc};1,1})
+\cdots
,
\end{eqnarray} 
where 
\begin{eqnarray}
{C}_{\text{gc};1,1} =
\tilde{\omega}^{5/2} \sum_{i=0}^{\infty} 
{c}_{\text{gc};1,1}^{(i)} \tilde{\omega}^{i} a_{\text{ho}}^{-1}
\end{eqnarray}
and 
\begin{eqnarray}
{C}_{\text{gc};2,1}
-2{C}_{\text{gc};1,1}
=
\tilde{\omega}^{5/2} \sum_{i=3}^{\infty} 
{c}_{\text{gc};2,1}^{(i)} \tilde{\omega}^{i} a_{\text{ho}}^{-1}.
\end{eqnarray}
The coefficients 
$c_{\text{gc};1,1}^{(i)}$ and $c_{\text{gc};2,1}^{(i)}$
are listed in 
Table~\ref{table_taylor2}.
\begin{table}
\caption{
  High-temperature expansion coefficients 
${c}_{\text{gc};1,1}^{(i)}$ and ${c}_{\text{gc};2,1}^{(i)}$
extracted from the contact determined in
the grand canonical ensemble.
}
\label{table_taylor2}
\centering
\begin{tabular}{c|cccc}
\hline
\hline
 $i$  &  0  & 1  
 &  2 & 3   \\
 \hline
 ${c}_{\text{gc};1,1}^{(i)}$  & $4\sqrt{\pi}$ & 0 &
 ${\sqrt{\pi}}/{6}$ & $-7.01342(2)$ \\
 ${c}_{\text{gc};2,1}^{(i)}$  &  &  &  & $-7.01342(2)$\\
\hline
\hline
\end{tabular}
\end{table} 
Comparison of Tables~\ref{table_taylor} and 
\ref{table_taylor2} shows that the coefficients with
$i=0-2$ agree but that discrepancies exist for $i=3$.
Specifically, our analysis shows 
that the leading order three-body coefficients
agree, i.e., $c_{\text{gc};2,1}^{(3)}=c_{2,1}^{(3)}$,
but that the two-body term at the same order in $\tilde{\omega}$
does not agree, i.e.,
$c_{\text{gc};1,1}^{(3)} \ne c_{1,1}^{(3)}$.
Disagreement is expected since the 
canonical and grand canonical ensembles are known to yield different 
results. The fact that the disagreement appears in the term proportional
to $N^2$ and not the term proportional to $N^3$ makes sense,
since our analysis in the grand canonical ensemble assumes large $N$, rendering
the term proportional to $N^2$ less important than the term proportional to
$N^3$.